\documentclass[manuscript]{aastex}

\usepackage{subfigure}

\slugcomment{submitted to {\it Astronomical Journal (v.0628.2016)}}

\shorttitle{Dynamical Masses of Subdwarfs}
\shortauthors{Jao et al.}

\begin{document}

\title{Cool Subdwarf Investigations III: Dynamical Masses of Low
  Metallicity Subdwarfs}

\author{Wei-Chun Jao\altaffilmark{1}}
\affil{Department of Physics and Astronomy, Georgia State University,
 Atlanta, GA 30302}
\email{jao@astro.gsu.edu}

\author{Ed. P. Nelan}
\affil{Space Telescope Science Institute, 3700 San
Martin Drive, Baltimore, MD 21218, USA}
\email{nelan@stsci.edu}

\author{Todd J. Henry\altaffilmark{1}}
\affil{RECONS Institute, Chambersburg, PA 17201}
\email{thenry@astro.gsu.edu}

\author{Otto G. Franz, Lawrence H. Wasserman}
\affil{Lowell Observatory, 1400 West Mars Hill Road, Flagstaff, AZ 86001}
\email{ogf@lowell.edu, lhw@lowell.edu}

\altaffiltext{1}{Visiting Astronomer, Cerro Tololo Inter-American
Observatory.  CTIO is operated by AURA, Inc.\ under contract to the
National Science Foundation.}

\begin{abstract}

We report dynamical mass measurements for the components of the
previously known double-lined spectroscopic subdwarfs G 006-026 B and
C using the Fine Guidance Sensors (FGS) on the {\it Hubble Space
  Telescope}. To build the empirical mass-luminosity relation for low
metallicity subdwarfs, we collect four other subdwarf systems with
dynamical masses that we compare to theoretical models for various
metallicities on the mass-luminosity relation. For most stars, they
fall in the regions where the models predict to be low
metallicity. This effort highlights the scarcity of dynamical masses
for subdwarfs and that much work remains to be done to improve the
mass errors and metallicity measurements of low mass subdwarfs in our
Galaxy.

\end{abstract}

\section{Introduction}
\label{sec:intro}

Subdwarfs are Galactic fossils that presumably comprise the bulk of
the halo, and are crucial touchstones of the star formation and metal
enrichment histories of the Milky Way.  The local paucity of subdwarfs
and their intrinsic faintness make them difficult to characterize,
unlike their disk counterparts.  For example, there are currently only
three confirmed subdwarf systems within 10 parsecs: $\mu$ Cas AB, CF
UMa, and Kapteyn's Star \citep{Monteiro2006}.  Many subdwarfs are
detected via high proper motion star surveys (e.g.~\citealt{Ryan1991},
\citealt{Carney1994}, \citealt{Gizis1997}, \citealt{Jao2005, Jao2008},
\citealt{Lepine2007}, and \citealt{Savcheva2014}), yielding important
information about their kinematics, chemical compositions, and space
densities.  However, their masses --- the single most important
parameter for a star --- are still mostly unknown, but would be of
great use in studies of the Galactic halo, globular clusters, and
other old and/or low metallicity systems.

There are currently only a half dozen subdwarf systems (here defined
as stars having [m/H] or [Fe/H]$\leq$$-$0.5) with dynamical mass
measurements. \cite{McCarthy1993} used infrared speckle interferometry
and \cite{Drummond1995} used Adaptive Optics to measure the masses of
$\mu$ Cas AB, which has [Fe/H]$=-$0.71 \citep{Karaali2003} and is one
of the nearest subdwarf systems.  \cite{Soderhjelm1999} reported
masses for GJ 704 AB (99 Her), which has [Fe/H]$=-$0.58 (See
Table~\ref{tbl:GJ704mH}).  Recently, \cite{Horch2015} reported total
masses for two subdwarf binaries (HIP 85209AB and HIP 95575AB) using
the Differential Speckle Survey Instrument (DSSI) to resolve these two
doubled-lined spectroscopic binaries \citep{Goldberg2002,
  Halbwachs2012}.  We use the mass-ratio from SB2 results and total
masses from \cite{Horch2015} to get individual masses. All of these
results are summarized in Table~\ref{tbl:mass}.  We note that HIP
81023AB and HIP 103987AB both have metallicity less than $-$0.5, but
have preliminary orbital results from \cite{Horch2015}. We did not
include them in our analysis. \cite{Ren2013} combined both
single-lined spectroscopic data, photocentric orbital data from the
Hipparcos Intermediate Astrometric Data and an empirical
mass-luminosity relation to estimate masses for four additional
subdwarf systems with [Fe/H]$\leq$$-$0.5 (HIP 705, HIP 39893, HIP
55022, HIP 59750 and HIP 73440).  However, these masses are not
measured dynamically, as they rely on an empirical mass-luminosity
relation from stars with solar metallicity, so these stars are not
included in our discussion. Finally, \cite{Soderhjelm1999} reported
the total system mass and mass ratio of GJ 60AB with [Fe/H]$=-$0.65
\citep{Holmberg2009}.  A few years later, \cite{Watson2001} reported
that the system could harbor four components rather than two, with
close binaries A-C and B-D, where B-D is an eclipsing binary with a
period of 0.45 days.  Because of the complexity of this system, we
conclude that no reliable masses for individual components have been
measured, so do not include this system in the discussion.

Nearby galactic globular clusers (GCs) are also homes of low
metallicities stars.  Recently, the Clusters AgeS Experiment (CASE)
project \citep[and references therein]{Kaluzny2015} has detected and
measured many detached eclipsing binaries in $\omega$ Cen, 47 Tuc, M4,
M55 and NGC6362.  These binaries are comprised of either massive or
evolved stars, and their metallicities are much lower than nearby
field subdwarfs, so they are not discussed further in this paper
either.

Because the local known subdwarf density is far less than that for
dwarfs, coupled with the fact that the multiplicity of cool subdwarfs
appears to be lower than that of main sequence stars \citep{Jao2009,
  Lodieu2009}, it has been difficult to identify appropriate binary
systems to target for dynamical mass determinations.  This is
particularly true for low mass M-type subdwarfs with masses less than
0.6 $M_{\odot}$.  Consequently, the number of dynamical masses
measured for low metallicity subdwarfs is much less than for their
counterparts, the main sequence dwarfs (\citealt{Henry1999,
  Delfosse2000, Torres2010, Benedict2016}).  Building an empirical
mass-luminosity relation for low metallicity stars has proven
difficult, and rigorous testing of low metallicity theoretical models
remains to be done.  In this manuscript, we present new dynamical
masses of $\sim$0.47 and 0.44 $M_{\odot}$ for the low metallicity
subdwarfs, G 006-026 B and C, adding two subdwarf masses to establish
the empirical mass-luminosity relation.  We also recalculate the
masses for GJ 704 A and B by reassessing the resolved measurements to
date.

G 006-026 BC ia a double-lined spectroscopic binary with period of
about 302 days \citep{Goldberg2002}.  \cite{Carney1994} reported that
the primary star in the system, the G-type star G 006-026 A, has a
metallicity of $[m/H]= -0.88$. Later, \citet{Casagrande2011} and
\cite{Holmberg2009} use photometry to determine the primary's
$[Fe/H]=-0.52$ and $-0.60$, respectively. We assume that the B and C
components have the same metallicity as the primary, and that they
therefore meet our criterion for subdwarfs as stars having
[Fe/H]$\leq$$-$0.5. Five of the seven orbital elements have been
determined: period ($P$), time of the periastron ($T$), semi-major
axis ($a$), eccentricity ($e$) and position of periastron ($\omega$).
The binary's orbital inclination ($i$) and longitude of the ascending
of node ($\Omega$) remain unknown from spectroscopic data
alone. Because of its relatively large $a\sin i$ value among the 34
SB2 systems in \citet{Goldberg2002}, we expected the Fine Guidance
Sensor on the Hubble Space Telescope to be able to resolve the system
at appropriate orbital phases. By accurately measuring the angular
separation of the two components at only a few epochs, we can reliably
determine all of the orbital elements and thus the individual masses
of the B and C components.

\begin{deluxetable}{lccccccccc}
\rotate
\tablewidth{0pt}
\tabletypesize{\small}
\tablecaption{Metallicities and Dynamical Masses of Subdwarfs\label{tbl:mass}}
\tablehead{
\colhead{Name}   &
\colhead{$\pi$}  &
\colhead{[m/H]}  &
\colhead{[Fe/H]} &
\colhead{Ref.}   &
\colhead{Mass (A)} &
\colhead{$M_{V}$}  &
\colhead{Mass (B)} &
\colhead{$M_{V}$}  &
\colhead{Ref.}
\\
\colhead{}       &
\colhead{mas}    &
\colhead{}       &
\colhead{}       &
\colhead{}       &
\colhead{$M_{\odot}$} &
\colhead{}       &
\colhead{$M_{\odot}$} &
\colhead{}       &
\colhead{}
}
\startdata
HIP 85209 AB   & 19.76$\pm$0.82 &($-$0.75)$^b$&          &  2  &  0.84$\pm$0.09      & 5.42 &  0.74$\pm$0.10      &  5.41 & 1$^a$ \\
HIP 95575 AB   & 39.98$\pm$0.73 & $-$0.80     &          &  3  &  0.69$\pm$0.10      & 6.80 &  0.58$\pm$0.09      &  7.03 & 1$^a$ \\
$\mu$ Cas AB   &132.67$\pm$0.74 &             &$-$0.71   &  9  &  0.742$\pm$0.059    & 5.78 &  0.173$\pm$0.011    & 11.6  & 4 \\
GJ 704 AB      & 64.30$\pm$0.68 &             &$-$0.58   &  6  &  0.89$\pm$0.03$^c$  & 4.16 &  0.51$\pm$ 0.03$^c$ &  7.46 & 5 \\
G 006-026 BC   & 25.67$\pm$0.04 & $-$0.88     &          &  8  &  0.474$\pm$0.053    &10.34 &  0.436$\pm$0.049    & 10.64 & 7 \\
\enddata 

\tablecomments{All parallaxes other than $\mu$ Cas AB and GJ704 AB are
  from \cite{Leeuwen2007}. The weighted mean parallax of $\mu$ Cas AB
  is from \cite{Leeuwen2007} and \cite{YPC}. The weighted mean
  parallax of GJ704AB is from \cite{YPC} and
  \cite{Soderhjelm1999}. $^a$\citet{Horch2015} did not report
  individual mass errors, but they have total mass ($M_{A}+M_{B}$)
  errors. We adopt these total mass errors to get errors in percentage
  of masses and apply them to each component.  $^b$The metallicity of
  HIP 85209 is discussed in $\S$6.  $^c$The masses of GJ 704 AB are
  discussed in $\S$\ref{sec:GJ704AB} and details are given in Table~\ref{tbl:GJ704}}

\tablerefs{
(1) \cite{Horch2015},
(2) \cite{Latham1992},
(3) \cite{Holmberg2009},
(4) \cite{Drummond1995},
(5) \cite{Soderhjelm1999},
(6) mean value of several measurements given in Table 5,
(7) this work,
(8) \cite{Carney1994},
(9) \cite{Karaali2003}
}

\end{deluxetable}

\section{Observations and Results}
\label{sec:obs}

We utilized one of {\it Hubble Space Telescope's} three Fine Guidance
Sensors (HST/FGS1r, hereafter simply FGS) in its TRANS mode to resolve
G 006-026 BC.  The FGS is a white light shearing interferometer. When
operated in it high angular resolution TRANS mode, it scans a luminous
source and produces interference fringes along two orthogonal axis (X
and Y). Resolved binary systems yield interference fringes, or
S-curves, that have reduced amplitude and different morphology than
those of point sources (unresolved stars). Model binary systems, built
from point source templates, are used to determine the projected
separation and relative bightness of the binary components along both
the FGS X and Y axis. We carried out TRANS mode observations of G
006-026 BC during {\it HST} Cycles 15 and 19 to measure the separations
and position angles for the C component relative to B at six epochs.
Typically, we scheduled 30--32 scans on both X and Y axes during each
visit, with scan lengths of 0\farcs4--1\farcs0 on either side of the
target.  Observations were made through the F583W filter, which
provides magnitude differences similar to the $V_J$ band in the
Johnson system (hereafter, simply $V$).

The spectroscopic data \citep{Goldberg2002}, available mass estimates,
and the Hipparcos distance to G 006-026 A indicated that the maximum
separation of the two stars would be $\pm$25 milli-arcseconds (mas).
However, because we did not know the position angle of the binary (due
to lack of knowledge of orbit's inclination and longitude of ascending
node), it was not possible to select with certainty an orientation of
the {\it HST} that would guarrentee that the binary would be resolved
along both axes of the FGS, which has a per axis angular resolution
close to 10 mas.  For the first observation in 2006, we selected an
observation date that corresponded to a time of maximum separation
assuming an inclination of 45 degrees, but still with no knowledge of
the longitude of ascending node. The binary was well resolved along
the FGS X-axis but unresolved along the Y-axis. The observed
separation along the X axis allowed for an estimate of the system's
inclination but left a position angle uncertainty of about +/- 27
degrees (because the actual Y-axis separation can be anywhere from
+/-10 mas). Subsequent observations in 2011 through 2014 were planned
using the spectroscopic orbital elements combined with the FGS data
previously acquired. For our last epoch in 2014, we observed the
binary in two consecutive {\it HST} orbits with a telescope roll
change of 41 degrees between them in an attempted to resolve the
system along both FGS axes, or to at least tightly constrain the
binary position angle.

The standard {\it strfits} routine in the IRAF/STSDAS package was used
to convert the archived TRANS mode FITS files into GEIS files (generic
edited information sets, the original file structure used for the
legacy {\it HST} instrument calibration pipelines, but still in use
for FGS). For each visit these GEIS data were processed through the
FGS raw data processing routine {\it calfgsa}, resulting in sets of
individual interference scans in the X and Y directions across the
binary. The analysis package {\it ptrans} was then used to
cross-correlate, and co-add the individual scans, and to smooth the
final co-added X and Y axis scans for each visit. Additional details
of the reduction procedure can be found in the {\it HST/FGS Data
  Handbook} (http://www.stsci.edu/hst/fgs/documents/datahandbook/).

The analysis of FGS TRANS mode observations of binary stars requires
the availabilty of TRANS mode observations of single (point source)
calibration stars of similar color. In January 2009 the FGS1r internal
optics were re-aligned to re-optimize (change) the instrument's
interferometric response. Thus our early 2006 epoch observation uses
SAO 185689 as the calibration star, which was observed in 2006, while
all later epochs use LHS 73, which was observed in 2009 after the FGS
re-optimization. The same data processing routines and packages
discussed above were applied to these calibaration stars. Routine
monitoring of FGS TRANS mode performance by STScI indicates that its
interference fringes have not changed since FGS's realignment in 2009,
so use of LHS 73 data from 2009 as the point source template for the
later epochs is justified.  We note that LHS 73 ($V-K_s=$3.44) is a
known low metallicity K6 subdwarf \citep{Jao2008} with a similar color
to G 006-026 BC ($V-K_s=$4.14). Two other often used FGS calibrators,
Latcol-A and HD233877, do not yield better fits/residuals than LHS 73,
so we use LHS 73 as our calibrator.

\begin{figure}
  \centering
  \includegraphics[angle=0, scale=0.50]{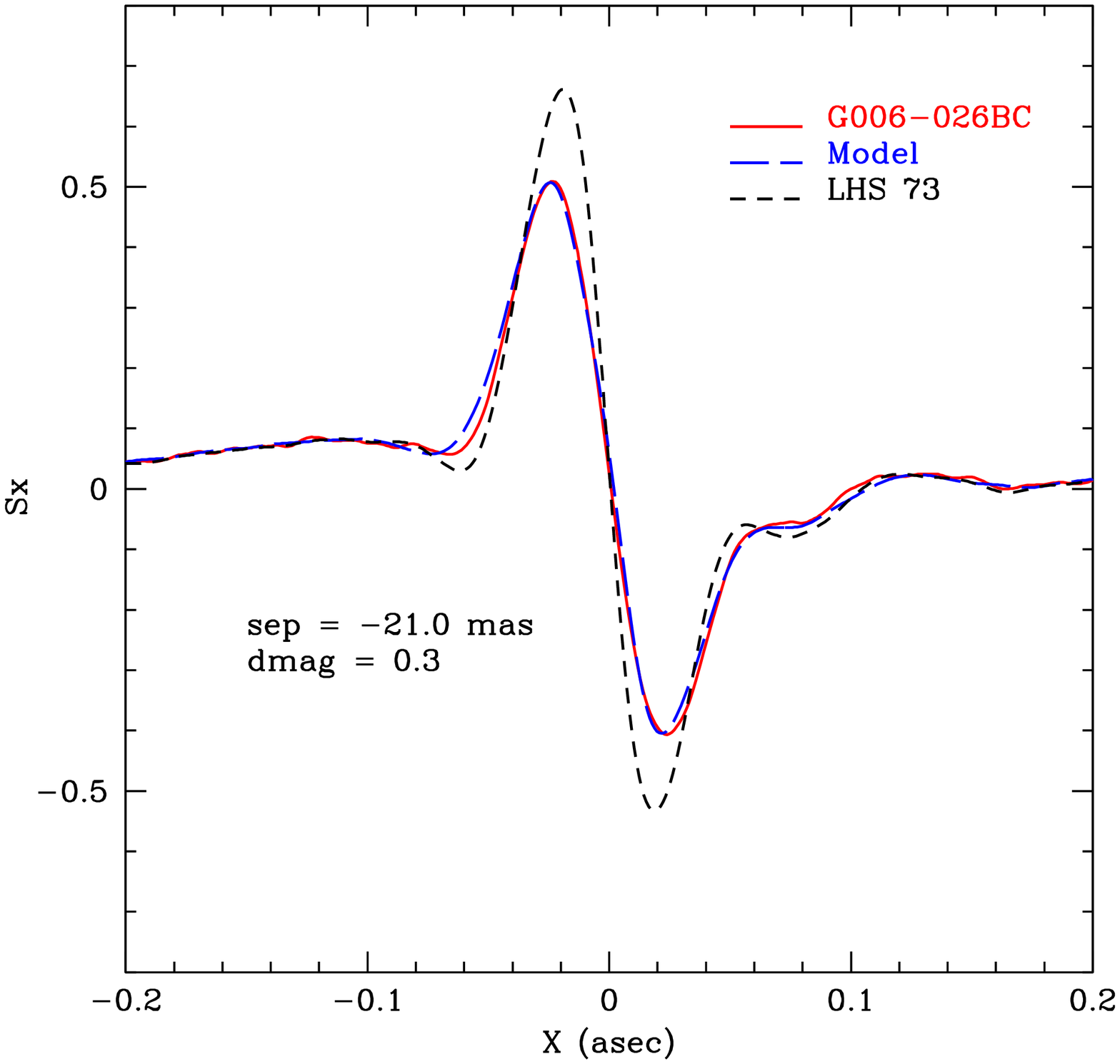}
  \includegraphics[angle=0, scale=0.50]{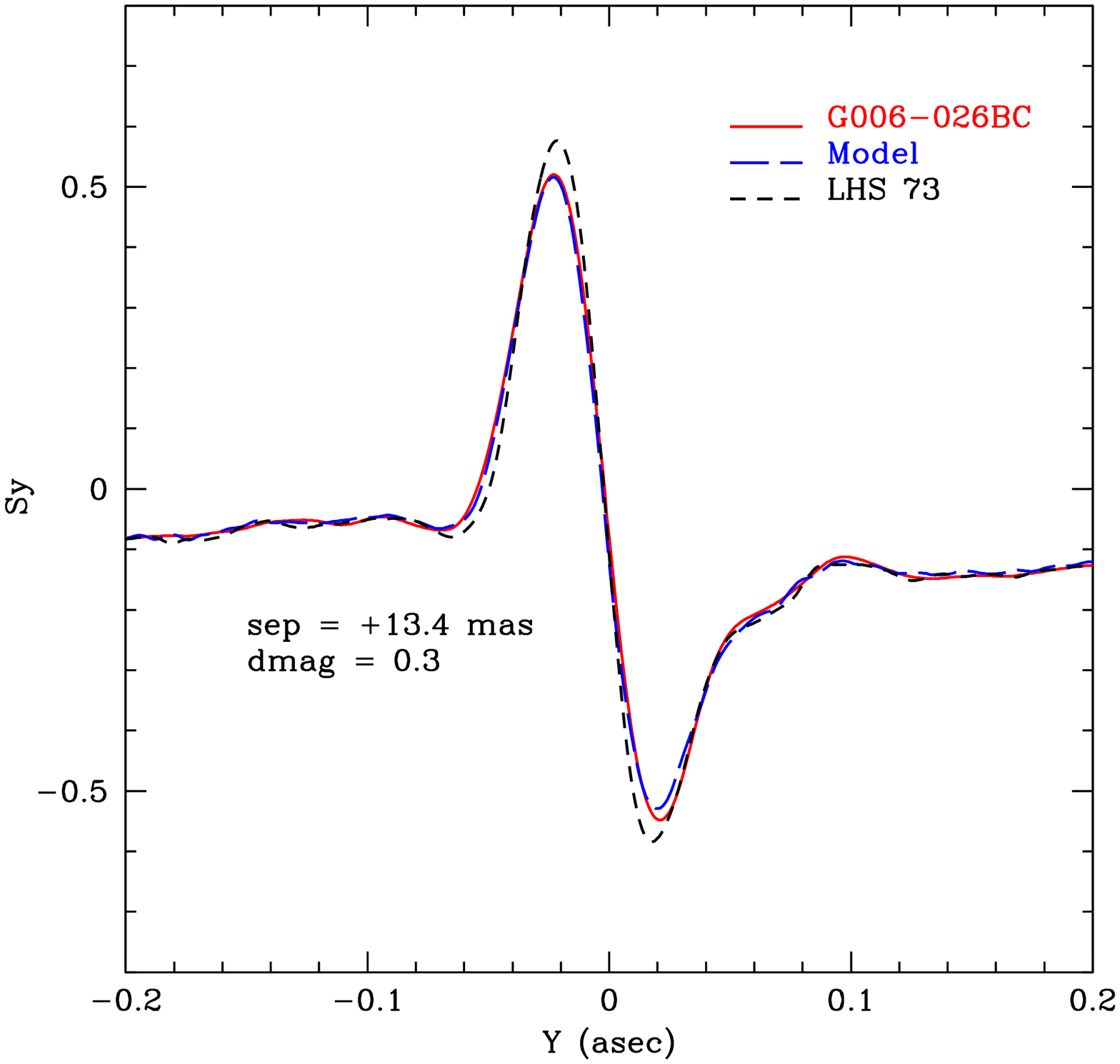}

  \caption{The FGS X and Y axis interference fringes for the
    observation of G 006-026 BC and the best fitting model are
    shown. For perspective, the interference fringes for the point
    source calibration star LHS 73 are also shown. The Y-axis
    separation of 13.4 mas approaches the 10 mas angular resolution
    limit of the FGS.}

\label{fig:fgsfit}
\end{figure}

\begin{figure}
\centering
\includegraphics[scale=0.6, angle=-90]{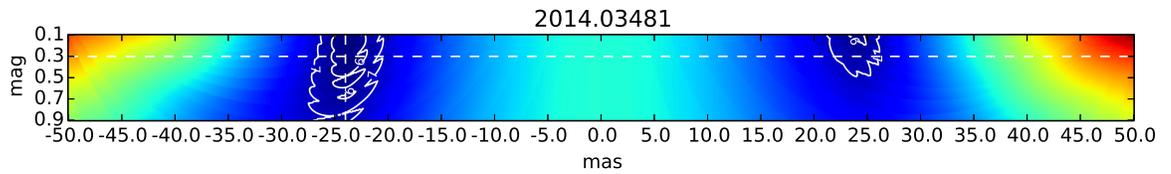}
\caption{This color plot shows the residuals of the fitting while
  changing the $\Delta$m and separation of the secondary in
  2014.03481.  Residuals rise from blue to red. Two white contours
  represent the equal residuals of the two lowest residuals 6 (inner
  contour) and 7 (outer contour). Two dashed lines indicate the
  location of the best fit from {\it binary\_fit} routine. }
\label{fig:2Dcolor}
\end{figure}

Finally, we used the standard {\it binary\_fit} package discussed in
the {\it HST FGS Data Handbook} to compare the interference fringes,
or S-curves, of G 006-026 BC to the calibrators to calculate the best
fitting component magnitude difference and separations in the X and Y
directions. Figure~\ref{fig:fgsfit} shows an example fitting result
for both axes in 2013.09. The {\it binary\_fit} program yields the
best $\Delta$m and separation by comparing with the
calibrator. However, we notice that the fitting for this system is not
sensitive to the $\Delta$m for all scans as shown in
Figure~\ref{fig:2Dcolor}. Because of this reason, we fix the $\Delta$m
at 0.3 with an error of 0.3 mag and apply this fixed value for all
other scans while we use {\it binary\_fit} routine.

We converted these separations along the FGS axes to separations
$\rho$ and position angles $\theta$ on the sky using the {\it HST}
keyword PA$\_$APER in the GEIS header, which is the position angle of
the FGS Y-axis at the time of the observation.  Results for the six
epochs of observation are given in Table~\ref{tbl:sepPA}. Measurement
errors in the $\rho$ and $\theta$ are shown in parenthess. For epochs
2006.55227 and 2011.83228 the binary was only resolved along one of
FGS axis, therefore the resultant uncertainty in position angle is
large (assuming up to 10 mas uncertainty in separation along the
unresolved axis). The data acquired in 2014 has the binary marginally
resolved along the FGS Y-axis. For such small separations, solutions
of the opposite parity must be considered (i.e., a separation of -12.4
mas is nearly indistinguishable from a separation of +12.4
mas). However, a different parity of the Y-axis separation for either
of the two visits produces incompatible position angles (205.26
degrees for the first visit, 299.63 degress for the second
visit). Thus by having changed the {\it HST} roll angle by 41 degrees
between the two visits, we are able to select the correct parity and
tightly constrain the binary position angle at this epoch.

\begin{deluxetable}{cccccc}
\centering
\tablecaption{HST/FGS TRANS mode results\label{tbl:sepPA}}

\tablehead{
\colhead{epoch}   &
\colhead{X sep}   &
\colhead{Y sep}   &
\colhead{PA Aper} &
\colhead{$\rho$}  &
\colhead{$\theta$} 
\\
\colhead{}        &
\colhead{mas}     &
\colhead{mas}     &
\colhead{deg}     &
\colhead{mas}     &
\colhead{deg}     
}
\startdata
2006.55227                   & +22.7  &  -7.0 & 170.6 & 23.8(2) & 277.7(27)  \\
2011.83228                   &  +8.9  &  18.5 & 170.2 & 20.6(2) & 195.9(33)  \\ 
2013.09010                   & -21.0  &  13.4 & 344.0 & 24.9(2) & 286.6(5)   \\ 
2013.12103                   & -20.8  &  13.1 & 337.0 & 24.6(2) & 279.3(5)   \\
2014.03481 \tablenotemark{*} & -21.7  &  12.4 & 325.0 & 24.8    & 265.0      \\ 
2014.03498 \tablenotemark{*} & -24.0  & -10.5 &   6.0 & 26.2    & 252.0      \\
\enddata

\tablenotetext{*}{We combine these two epochs of data to get a mean
  separation (25.5$\pm$2 mas) and position (258.5$\pm$7$^\circ$) angle
  in 2014.0349 for our orbital fitting routine. The individual errors
  of $\rho$ and $\theta$ of these two epochs are not shown in this table.  }

\end{deluxetable}

\section{G 006-026 BC}
\label{sec:G006-026BC}

\begin{figure}
\centering
\includegraphics[scale=0.6, angle=90]{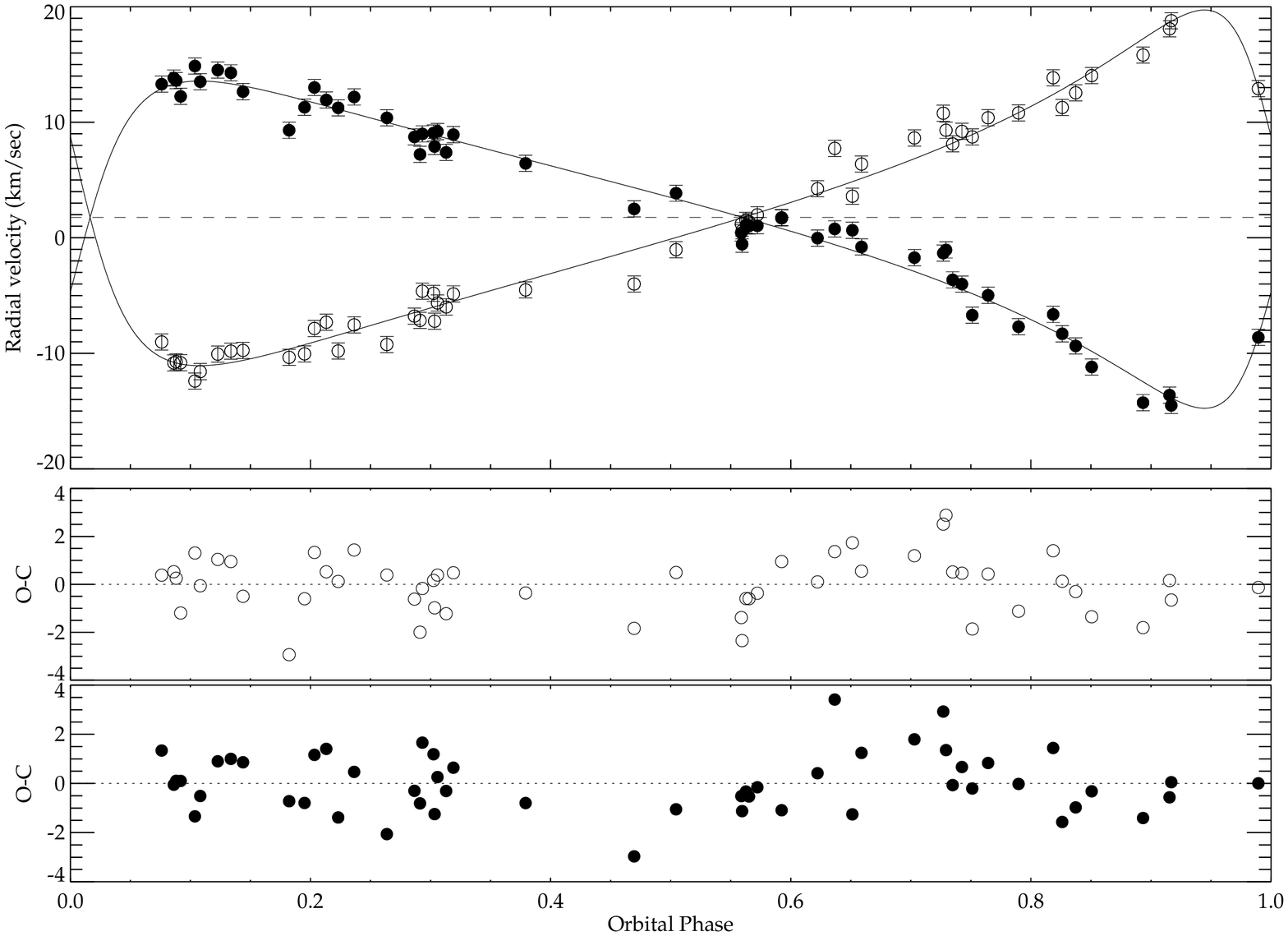}
\caption{The radial velocity data for G 006-026 BC from
  \citet{Goldberg2002}.  Filled and open circles are for the primary
  and secondary velocities, respectively.  Residuals are shown on the
  bottom of the figure for both primary and secondary. Because there
  is no error available for individual radial velocity measurements,
  the mean error of 0.7 km/sec is uses for all measurments.}
\label{fig:G006SB}
\end{figure}

\begin{figure}
\centering
\includegraphics[scale=0.6, angle=90]{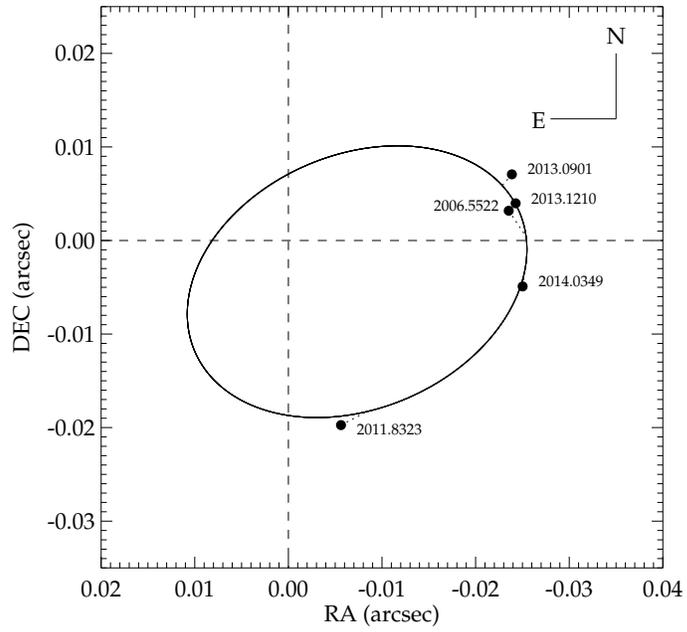}
\caption{The relative orbital motion of G 006-026 BC is shown using
  the FGS results.  Dashed lines connect observed and calculated
  positions.}
\label{fig:G006VB}
\end{figure}

We used the MIIPS (Multipurpose Interactive Image Processing System,
\citep{Gudehus2001}) package to calculate all seven orbital elements
and as well as dynamical masses for the B and C components by using
the FGS results from this work and radial velocities from
\citet{Goldberg2002}.  The radial velocity dataset includes 43
epochs/observations taken with the Center for Astrophysics Digital
Speedometers, which provide velocities with residuals on the order of
1 km/s.  The radival velocity data are shown in
Figure~\ref{fig:G006SB}, with the fit to the data from our work.  A
total of five epochs of astrometric data from the FGS is used to
complete the determination of the system’s orbital
elements. Figure~\ref{fig:G006VB} shows the plot of G 006-026 BC's
relative orbit. Table~\ref{tbl:G006} shows our fit to the orbital
elements along with those from \cite{Goldberg2002}.

\begin{deluxetable}{lcc}
\tabletypesize{\small}
\tablewidth{510pt}
\centering
\tablecaption{Orbital elements of G 006-026 BC\label{tbl:G006}}
\tablehead{
\colhead{Parameters}    &
\colhead{Goldberg}        &
\colhead{this work}
}
\startdata
Period ($P$~yr)                                  &  0.826$\pm$0.001     & 0.826$\pm$0.0003\\
Semi-major axis ($a$~$\arcsec$)                  &  \nodata             & 0.0108$\pm$0.0004\\
Eccentricity ($e$)                               &  0.582$\pm$0.017     & 0.577$\pm$0.004 \\
Inclination ($i$~$^\circ$)                        &  \nodata             & 128.31$\pm$4.98\\
Ascending Node ($\Omega$~$^\circ$)                &  \nodata             & 123.12$\pm$3.42\\
Longitude of Periastron ($\omega$~$^\circ$)       &  252.8$\pm$1.9       & 253.24$\pm$0.46\\
Epoch of Periastron ($T$~yr)                      & 1991.4385$\pm$0.004 &1991.4394$\pm$0.0008\\
System Velocity ($\gamma$~km sec$^{-1}$)          & 1.76$\pm$0.12        &1.76$\pm$0.03 \\
Primary Amplitude ($K1$~km sec$^{-1}$)            & 14.25$\pm$0.42       & 14.16$\pm$1.44\\
Primary Amplitude ($K2$~km sec$^{-1}$)            & 15.48$\pm$0.44       & 15.39$\pm$1.57\\
Parallax ($\pi$~$\arcsec$)                       & \nodata              &0.02634$\pm$0.0002\\
Total Mass (M$_{\odot}$)                          & \nodata              &0.911$\pm$0.110\\
M$_{A\odot}$                                      & \nodata              &0.474$\pm$0.053\\
M$_{B\odot}$                                      & \nodata               &0.436$\pm$0.049\\
\enddata
\end{deluxetable}

There are four ($P$, $e$, $\omega$, $T$) orbital elements and $\gamma$
(system velocity) from this work in common with measurements from
\citet{Goldberg2002}, and all are generally consistent.  However,
because this system is a challenge to resolve with FGS, the error in
orbital inclination (4.98$^\circ$) is relatively large and results in
larger mass errors (11\%) than anticipated.  By combining the visual
and spectroscopic orbits, we measure a high-quality orbital parallax
of 0\farcs02634$\pm$0\farcs0002, placing the system at 37.9 pc.  This
is consistent with the trigonometric parallax measured by {\it
  Hipparcos} for the primary, G 006-026 A,
0\farcs02561$\pm$0\farcs00134 \citep{Leeuwen2007}.


\section{GJ 704 AB}
\label{sec:GJ704AB}

GJ 704 AB is a visual binary with a $\sim$56 year orbital period at
15.6 pc.  The first visual measurement of the system dates back to
1859 \citep{WDS}.  \citet{Heintz1972}, \citet{Soderhjelm1999}, and
\citet{Kennedy2012} have all calculated orbits and individual masses
for the components, while \citet{Malkov2012} reported a total mass for
the system.  The system is also a single-lined spectroscopic binary
(SB1), with an orbit published by \citet{Abt2006}.  Orbital results
for all five studies are given in Table~\ref{tbl:GJ704}.

\begin{figure}
\centering
\includegraphics[scale=0.6, angle=90]{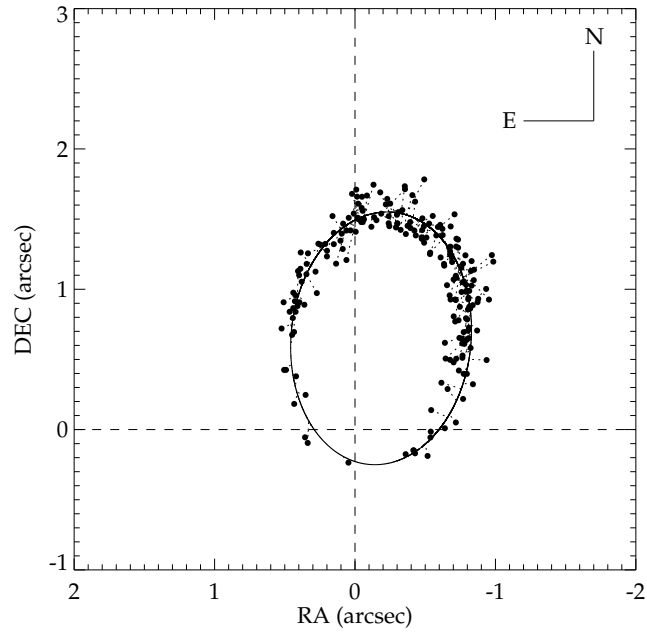}
\caption{The relative orbital motion of GJ 704 AB is shown using data
  from the WDS.  Dashed lines connect observed and calculated
  positions.}
\label{fig:GJ704VB}
\end{figure}

As of 2015, the Washington Double Star (WDS) catalog \citep{WDS} lists
240 different epochs of observations for GJ 704 AB, with the most
recent measurement being that of \citet{Kennedy2012}.  After removing
epochs with uncertain measurements and large errors, we have adopted a
dataset including 204 total measurements spanning 1859--2011 (i.e the
same coverage as \cite{Kennedy2012}), and recalculate the relative
orbit by combining both WDS resolved measurements and the radial
velocity data from \cite{Abt2006}.  For a given WDS epoch, we assign a
weight for the astrometric position based on telescope aperture,
system separation, and/or magnitude difference, following the
prescription of \cite{Hartkopf2001}.  Table~\ref{tbl:GJ704} lists our
new set of orbital elements, which are consistent with previous
efforts.  Our resulting orbital fit is shown in
Figure~\ref{fig:GJ704VB}, where it is clear that there are few
reliable measurements near periastron.  We use the period and
semi-major axis of this orbit and the weighted mean parallax of 64.3
mas from \cite{YPC} and \cite{Soderhjelm1999} to derive a total mass
for the system of 1.40 $\pm$ 0.03 M$_{\odot}$.

With the total system mass and orbital elements in hand, there are two
ways to calculate the masses of the components.  First, we use the
mass function ($\it f(m)=$0.01273) of the SB1 reported by
\citep{Abt2006} to determine individual masses of 0.91 M$_{\odot}$ and
0.48 M$_{\odot}$.  Second, we use $\Delta V=$ 3.3 mag and the
photocentric semi-major axis of 0\farcs353 from \citet{Heintz1972} to
calculate the scale of the photocentric/relative orbit that yields the
mass fraction, {\it f or $M_{B}/(M_{A}+M_{B})$} = 0.379.  Given the
total mass, we then find masses of 0.86 M$_{\odot}$ and 0.53
M$_{\odot}$ for the components.  For the final masses, we adopt the
means of these two methods, and find masses of 0.89 M$_{\odot}$ and
0.51 M$_{\odot}$. The mean difference between these mean values,
0.03M$_{\odot}$, is adoped as their errors, as listed in
Table~\ref{tbl:GJ704}.

\begin{deluxetable}{lcccccc}
\tabletypesize{\tiny}
\tablewidth{510pt}
\centering
\tablecaption{Orbital Elements for GJ 704 AB\label{tbl:GJ704}}
\tablehead{
\colhead{Parameters}             &
\colhead{Heintz (1972)}          &
\colhead{S{\"o}derhjelm (1999)}  &
\colhead{Malkov (2012)}          &
\colhead{Kennedy (2012)}         &
\colhead{Abt (2006)}             &
\colhead{this work}
}
\startdata
Period (P~yr)                                  &  55.8    &   56.4     &   56.04  &  56.3     &  56.4(fixed)  & 55.91$\pm$0.12   \\
Semi-major axis (a~$\arcsec$)                  &  1.00    &   1.00     &   1.123  &  1.06     & \nodata       & 1.05$\pm$0.02  \\
Eccentricity (e)                               &  0.74    &   0.75     &   0.798  &  0.766    &  0.75(fixed)  & 0.761$\pm$0.007   \\
Inclination (i~$^\circ$)                       &  32.0    &   34.0     &  \nodata &  39       & \nodata       & 36.18$\pm$1.72  \\
Ascending Node ($\Omega$~$^\circ$)             &  218.7   &  216.0     &    \nodata &  41       & \nodata       & 223.62$\pm$3.21  \\
Longitude of Periastron ($\omega$~$^\circ$)    &  300.6   &  301.0     &  \nodata &  116      &  120.4        & 295.38$\pm$2.84  \\
Epoch of Periastron (T~yr)                     & 1941.8   &  1998.0    &  \nodata &  1997.62  &  1999.52      & 1997.8$\pm$0.09 \\
Parallax ($\pi$~$\arcsec$)                     & 0.064    &  0.0639    &  0.06393 &  0.06393  & \nodata       & 0.0643$\pm$0.00068 \\
Total Mass (M$_{\odot}$)                       &  1.49    &   1.25     &  1.73    &  1.4      & \nodata       & 1.40$\pm$0.03  \\
Astrometric Coverage (yr)                      &1859--1968&1859?--1997? &  unknown      &  1859--2011 &1988--1996 ($V_{rad}$)  &1859--2011  \\     
M$_{A\odot}$                                   & 0.90     & (0.75)     &  \nodata &  0.94     & \nodata       & 0.89           \\
M$_{B\odot}$                                   & 0.59     & (0.50)     &  \nodata &  0.46     & \nodata       & 0.51           \\
\enddata

\tablecomments{The span of astrometric coverage is for visual relative
  positions, $rho$ and $\theta$, except in the case of
  \citet{Abt2006}, where the duration of the radial velocity
  measurements is given.  The astrometric coverage for
  \citet{Soderhjelm1999} is estimated to be up to 1997, which is the
  latest astrometric reference in that paper.  We use $f$ = 0.4 and
  the total mass from \citet{Soderhjelm1999} to estimate individual
  masses for that effort.  \citet{Malkov2012} do not clearly state the
  timespan of observations used for their analysis. We note that the
  total masses are different between \cite{Heintz1972} and
  \cite{Soderhjelm1999}, even though their orbital elements are
  essentially identical. The values in this table are directly from
  their papers. Based on Kepler 3$^{rd}$ law, the total masses should
  be 1.49 and 1.25, respectively for \cite{Heintz1972} and
  \cite{Soderhjelm1999}. }

\end{deluxetable}

According to SIMBAD, fifty-five publications report abundances for GJ
704 AB since 1960.  Determining which [Fe/H] values are best is a
monumental task beyond the scope of this manuscript.  We select four
measurements of [Fe/H] reported in refereed publications during the
past 10 years and [Fe/H] measurements, including both spectroscopic
and photometric methods.  These measurements are given in
Table~\ref{tbl:GJ704mH}, where we list a mean value of [Fe/H] =
$-$0.58\footnote{\cite{Latham2016} similarly reports that the [Fe/H]
of GJ 704 AB is $-$0.59 according to their unpublished data.}  that
indicates this to be a low-metallicity binary.  We note only
\cite{Holmberg2009} used photometry to measure [Fe/H].

\begin{deluxetable}{lc}
\tablewidth{300pt}
\centering
\tablecaption{Metallicity Values for GJ 704 AB\label{tbl:GJ704mH}}
\tablehead{
\colhead{Publications} &
\colhead{[Fe/H]}
}
\startdata
\citet{Takeda2013}     &     $-$0.58 \\
\citet{Maldonado2012}  &     $-$0.6  \\
\citet{Korotin2011}    &     $-$0.6  \\
\citet{Holmberg2009}   &     $-$0.55 \\
\hline
mean                   &     $-$0.58
\enddata
\end{deluxetable}

\section{M$_V$ Values as Proxies for Luminosities}
\label{sec:lum}

To develop an empirical mass-luminosity relation for subdwarfs we use
$M_V$ in the Johnson system as a proxy for luminosity.  Two systems,
$\mu$ Cas AB and GJ 704 AB, have combined $V$, $\Delta V$, and
parallax measurements, so it is straightforward to obtain individual
$V$ magnitudes for each component.  However, the two subdwarf systems
from \cite{Horch2015} have $\Delta m_{692nm}$ and $\Delta m_{880nm}$
rather than $\Delta V$ measurements.  To convert $\Delta m_{692}$ from
this special narrow band to the $V$ filter, we may (1) use synthetic
spectra for low metallicity stars convolved with filter passbands to
convert from $\Delta m_{692nm}$ to $\Delta V$, or (2) build an
empirical relation from existing systems with both $\Delta m_{692}$
and $\Delta V$ measurements.  The first method requires knowing the
metallicity of each system and the spectral types or effective
temperatures of each component so that we can select corresponding
synthetic spectra with the same $[m/H]$ and $T_{eff}$.  However, in
order to make a direct comparison between empirical and model
mass-luminosity relations, here we try to limit the use of synthetic
spectra to derive any stellar parameters, including $V$, so we use the
second method.

\begin{figure}
\centering
\includegraphics[scale=0.8]{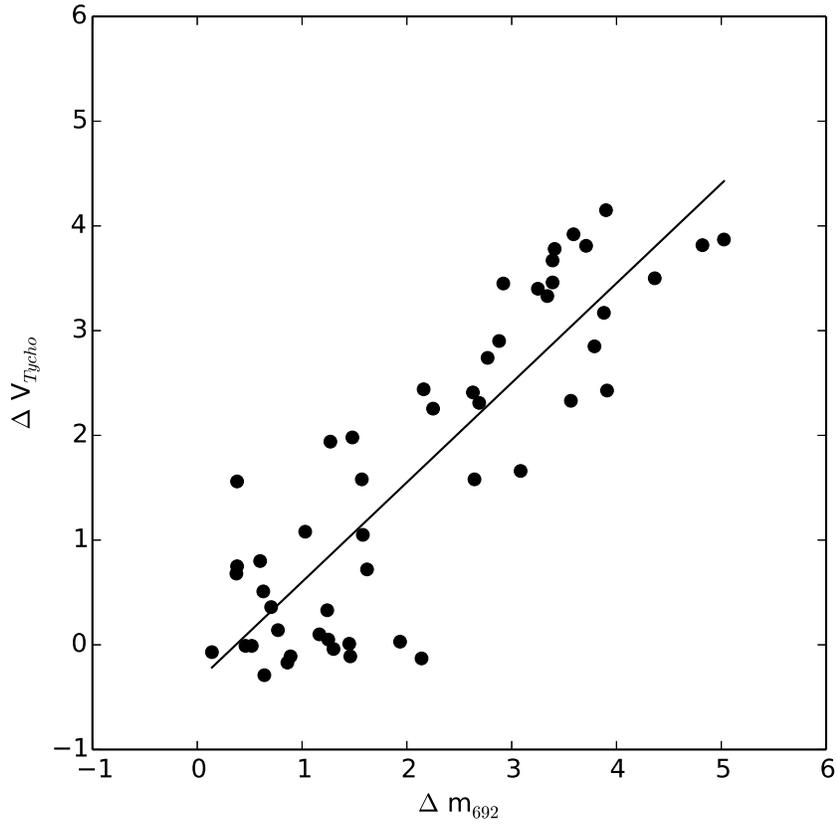}
\caption{Relation $\Delta V_{Tycho}$ vs $\Delta m_{692}$ from 51
  binaries in Washington Double Star Catalog. The solid straight line
  is a fit ($\Delta V_{Tycho}=-0.35+0.95\times\Delta m_{692}$) for these points.}

\label{fig:692totychov}
\end{figure}

To create a reliable conversion from $\Delta m_{692}$ to $\Delta V$,
we first searched the WDS for systems with both measurements.
Unfortunately, we found no such systems, but we did find 51 systems
with both $\Delta m_{692}$ and $\Delta V_{Tycho}$ measurements.  A
linear relation between measurements through these two filters is
shown in Figure~\ref{fig:692totychov}.

Ideally, we would then convert $\Delta V_{Tycho}$ to $\Delta V$ using
the polynomial conversion provided by \citep{Mamajek2002}.  However,
this conversion requires the $B_{T}-V_{T}$ colors of each star, which
is currently unavailable for any of two systems.  Nonetheless, we note
that the error in our relation shown in Figure 4 is $\sim$0.46 mag
between $\Delta V_{Tycho}$ and $\Delta m_{692}$, which is much larger
than the conversion error between Johnson $V$ and $V_{Tycho}$.
Therefore, until direct $\Delta V$ measurements are available, we
assume $\Delta V_{Tycho} \approx \Delta V$ for the two subdwarf
systems in \cite{Horch2015} and adopt $\sim$0.46 mag for our $V$
errors.

The G 006-026 BC system targeted with FGS also has no $\Delta V$
measurement, but we can use our $\Delta m_{583W}$ value from the FGS
observations discussed in section~\ref{sec:obs}.
\cite{Henry1999} have noted that the F583W filter in HST's FGSs nearly
matches the Johnson $V$ passband and provided a relation to convert
$\Delta m_{583W}$ to $\Delta V$.  Although their relation was derived
for stars presumed to have solar metallicity, we assume that even if
the {\it magnitudes} will be different for subdwarfs vs.~main sequence
stars, the {\it colors} between two stars in a system will not be
drastically different.  So, we assume $\Delta m_{583W} \approx \Delta
V$ for G 006-026 BC.

\section{The Empirical Mass-Luminosity Relation and Discussion}
\label{sec:MLR}

\begin{figure}
\centering
\includegraphics[scale=0.7, angle=90]{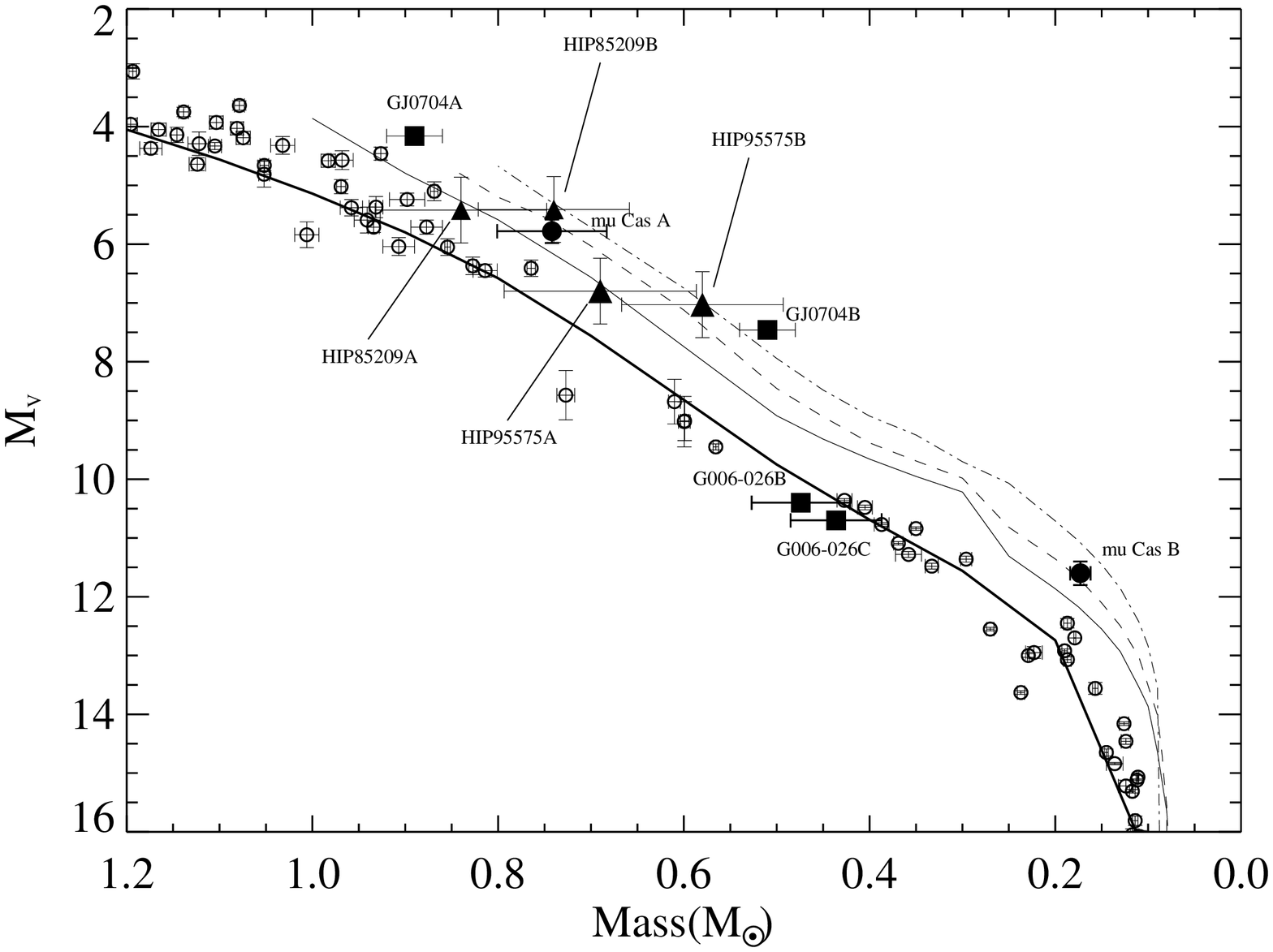}

\caption{The mass-luminosity relation for subdwarfs (filled symbols)
  and dwarfs (open circles).  Names are given for subdwarfs.  Open
  circles are from \cite{Delfosse2000}, \cite{Torres2010}, and
  \cite{Benedict2016}.  Subdwarfs from \cite{Horch2015} are filled
  triangles, with a different sizes used simply to match the two stars
  in a system.  Filled circles represent $\mu$ Cas AB and filled boxes
  represent the two systems discussed in detail in this paper, GJ 704
  AB and G 006-026 BC. The thick solid line represents the
  main-sequence isochrone for age 1 Gyr from BT-settl
  \citep{Baraffe2015}.  The three other lines represent the predicted
  locations of stars with metallicities of [m/H]=$-$0.5, $-$1.0 and
  $-$2.0 at an age of 6 Gyr from bottom to top, taken from
  \cite{Baraffe2015}}
\label{fig:mlr}
\end{figure}

We plot the 10 low-metallicity subdwarfs discussed here as solid
points on and empirical mass-luminosity relation shown in
Figure~\ref{fig:mlr}.  The masses come directly from Tables 1, 3 and
4, where in addition to the subdwarfs, main sequence stars with masses
from \cite{Delfosse2000}, \cite{Torres2010}, and \cite{Benedict2016}
are plotted in Figure~\ref{fig:mlr}.  These main-sequence stars
generally fall along the 1 Gyr theoretical isochrone from
\cite{Baraffe2015}, shown with a solid line.

The ages of low metallicity stars are primarily measured using nearby
GCs.  Recently, \cite{Bono2010} and \cite{Correnti2016} acquired deep
optical and near-IR photometry from the ground and HST to extend the
detected GC populations down to early M dwarfs.  Many of these
clusters/cool dwarfs have ages estimated to be around $\sim$11 Gyrs.
In addition, \cite{Monteiro2006} measured the first ages for two cool
field K subdwarfs via their white dwarf companions' cooling curves,
and found that these likely thick disk subdwarfs have ages of 6--9
Gyr.

However, until we can link the origins of these field subdwarfs to
nearby GCs, we conservatively adopt 6 Gyr as the age for the all
nearby subdwarfs.  We sketch 6 Gyr isochrones for stars with [m/H] =
$-$0.5, $-$1.0, and $-$2.0 from BT-Settl isochrones
\citep{Baraffe2015} in Figure~\ref{fig:mlr} to permit comparisons to
the empirical points for the subwarfs.  According to the models, low
metallicity subdwarfs appear brighter in $M_{V}$ than dwarfs at a
given mass, or alternately, are less massive for a given $M_{V}$.

Systems like $\mu$ Cas AB and GJ 704 AB are clearly seen above the
mass-luminosity relation of main sequence stars.  The locations of
$\mu$ Cas A and B, which have [m/H] = $-$0.71, are well-matched to the
model predictions for stars with [m/H] = $-$1.0 at an age of 6 Gyr.
The metallicity of GJ 704 AB appears much lower ([m/H]$<-$1.0) than
what has been measured [Fe/H]=$-$0.58, but the two subdwarfs certainly
appear to be of lower metallicity than main sequence stars.  We note
that \citet{Kennedy2012} used {\it Herschel's} Photodetector and Array
Camera and Spectrometer (PACS) at 100 and 160 $\mu$m to reveal a rare
and convincing circumbinary polar-ring debris disk around GJ 704 AB.
This makes this system very unusual because this disk has somehow been
maintained or created around a presumably old subdwarf binary.  
$M_V$ values used in Figure~\ref{fig:mlr} for GJ 704 A and B do not
include any adjustment for obscuration by the disk, which is likely to
be minimal.  If such an adjustment is required, the points would move
to brighter $M_V$, and the locations of the points would be towards
even lower metallicity.

The remaining subdwarfs systems generally fall along the low
metallicity curves. Because of the large mass errors of HIP 85209 AB
and HIP 95575 AB measured by \cite{Horch2015}, it is difficult to make
a direct comparison between observations and models. \cite{Latham1992}
reported a metallicity for HIP 85209 AB (HD 157948 AB) of
[m/H]=$-$0.75, whereas \cite{Goldberg2002} reported
[m/H]=$-$0.5. \cite{Horch2006} measured the dynamical masses of these
subdwarfs using FGS, and determined via comparisons to the theoretical
Yale-Yonsei models that the system should have metallicity close to
$-$0.5.  We note that the mass errors for HIP 85209 A and B are
relatively large, so these two stars do not place strict constraints
on the models.

Finally, both components in the new system described in this paper, G
006-026 B and C, merge with points for main sequence stars with solar
metallicity in the MLR.  This presents a conflict with the low
metallicity measurements discussed in Section~\ref{sec:intro} --- the
points should lie near the [m/H] = $-$0.5 or $-$1.0 lines in Figure 7.
For example, \cite{Carney1994} used high S/N echelle spectrograph to
determine the metallicity of the primary in the system G6-26A, which
is also a SB2 binary with a period of 8.7 days \citep{Goldberg2002}.
They discussed the challenges of measuring the metallicities of SB2
binaries in detail, but there is no flag of the system regarding
quality of their result of [m/H] = $-$0.88.  \citet{Casagrande2011}
and \cite{Holmberg2009} used photometry of the unresolved primary to
determine [Fe/H] = $-$0.52 and $-$0.60, respectively, thereby
providing results that are match Carney's, again indicating that the
system is metal-deficient.

\begin{figure}
\centering
\includegraphics[scale=0.7]{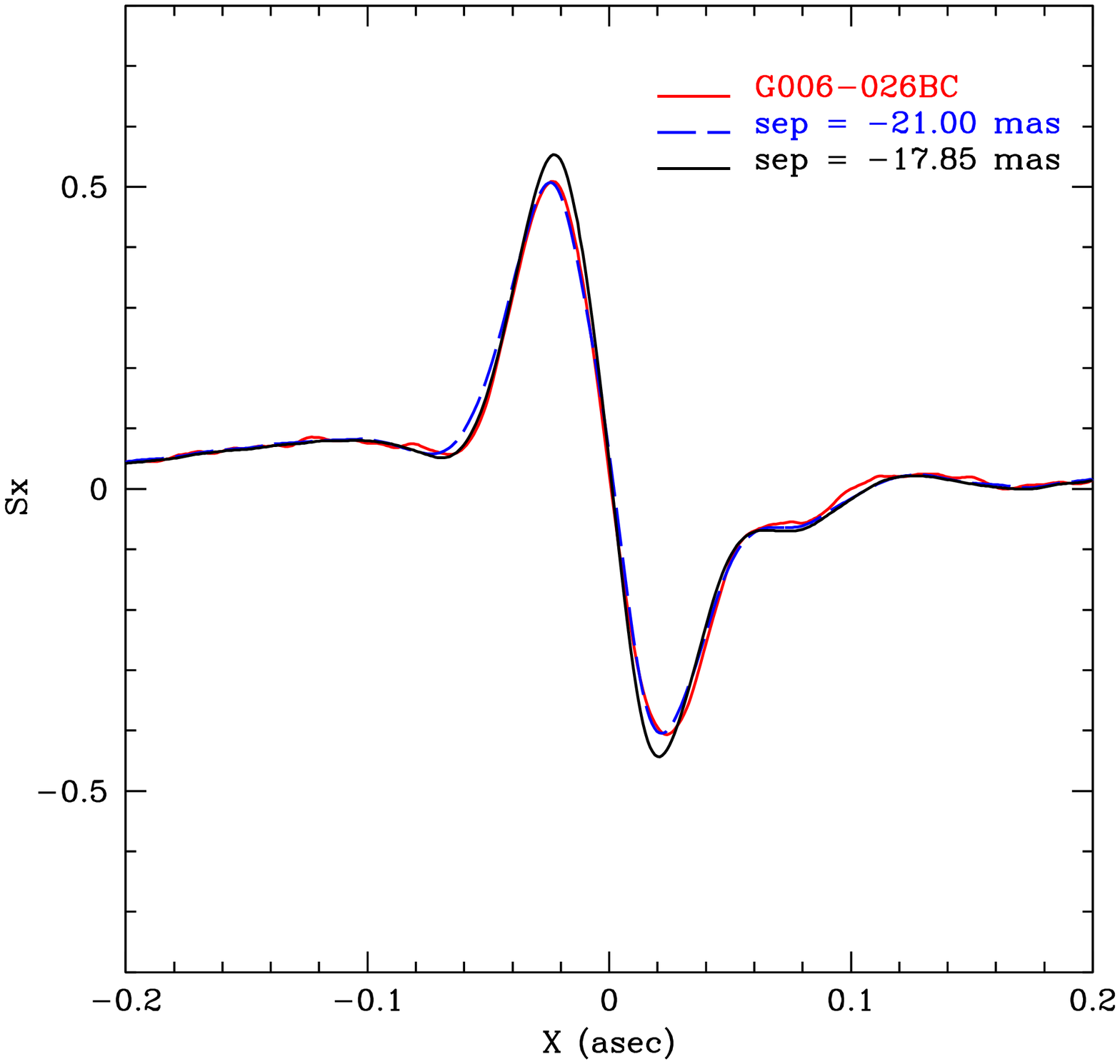}
\caption{Displayed is the best fitting model (blue line) to the
  2013.09 observation of G 006-26 BC (red line) along the FGS X-axis,
  with an angular separation of -21.0 mas. Also displayed is a model
  with an angular separation = -17.85 mas (black line), which would
  have resulted in the B component mass of $\sim$0.3 M$_{\odot}$ on
  the -0.5 theoretical metallicity line of the mass luminosity
  relation. Clearly this is inconsistent with the observation.}

\label{fig:fakeS}
\end{figure}

In order to move the masses of B and C components to the low
metallicity region, B component's mass would need to be reduced to
$\sim$0.3 M$_{\odot}$ (62.7\% of 0.478 M$_{\odot}$) to fall on the
$-0.5$ metallicity curve on Figure~\ref{fig:mlr}.  In Kepler's Third
Law, such a mass decrease requires that the semi-major axis drop by
15\%.  There would then be a corresponding reduction in the projected
separation ($\rho$) measured using FGS.  Figure~\ref{fig:fakeS} shows
an example of changing the projected separation along the X-axis by
15\% while comparing to our best fitted S-curve.  Note the poor fit to
the FGS data, compared to the much better correct fit.  In addition,
because our orbital parallax matches that from Hipparcos, it is
difficult to change the semi-major axis of this system from our
derived value of 10.8 mas.  Given that it is hard to dispute the three
independent metallicty measurements or shrink the separation between B
and C, we are left with no easy answer to the conundrum that that two
points do not fall on the MLR where they should.

Given the overall trend that subdwarf systems are elevated in the
mass-luminosity relation and match with the model, but there are
several confounding factors that challenge observational programs
focusing on mapping the mass-luminosity relation for subdwarfs,
including (1) metallicity errors may be large because all available
subdwarf binaries have separations less than 1\arcsec, so the acquired
spectra used to determine metallicities are for two objects, not one,
with consequent complications when measuring linewidths (G 006-026 AB
is the only system to have an early type primary, which can be used to
determine metallicity independently.), (2) mass errors are large
because subdwarfs are rare and further away than comparable main
sequence systems in the solar neighborhood, so parallax errors are
larger, (3) accurate $\Delta V$ measurements are not available for a
few systems and conversions from other filters are imperfect (although
overall, this is not likely to be a major problem), (4) not all of the
systems are of the same age, so comparison to a single isochrone for 6
Gyr-old systems may not be appropriate.

This work shows that building an empirical mass-luminosity relation
for subdwarfs has just begun.  Only two subdwarf systems, $\mu Cas$ AB
and GJ 704 AB, are clearly off the metallicity curves for
main-sequence stars and have low mass errors, so there remains
significant work to do in understanding the astrophysics that drives
the locations of low metallicity stars on the mass-luminosity
relation.  We can make progress in our understanding of these Galactic
fossils, as well as the star formation history and mass distribution
of stars in the Milky Way, by discovering more subdwarf binaries for
which high-quality metallicities, luminosities, and masses can be
measured.

\section{Acknowledgments}

We thank D. Latham and A. Bieryla to provide us their unpublished
metallicity of GJ 704 AB. We also thank the anonymous referee to
provide us very valuable comments to improve this manuscript.  The
HST-FGS observations were supported for program number 10927 and 12561
by NASA through grants from the Space Telescope Science Institute,
which is operated by the Association of Universities for Research in
Astronomy, Inc., under NASA contract NAS5-26555.

This research has made use of the SIMBAD database, operated at CDS,
Strasbourg, France.  This work also has used data products from the
Two Micron All Sky Survey, which is a joint project of the University
of Massachusetts and the Infrared Processing and Analysis Center at
California Institute of Technology funded by NASA and NSF. This
research has made use of the Washington Double Star Catalog maintained
at the U.S. Naval Observatory.



\end{document}